\documentstyle[12pt,epsf]{article}

\def\s{\sigma}

\def\be{\begin{equation}}
\def\ee{\end{equation}}
\def\o{\omega}
\def\bea{\begin{eqnarray}}
\def\eea{\end{eqnarray}}

\author{N. Sh. Izmailian\dag\S,\enskip Chin-Kun Hu\dag \enskip and
\enskip F. Y. Wu\ddag \\ \\
\dag\ {\normalsize Institute of Physics, Academia Sinica,
Nankang, Taipei 11529, Taiwan}\\
\S\ {\normalsize Yerevan Physics Institute, Alikhanian Br. 2, 375036
Yerevan, Armenia}\\
\ddag\ {\normalsize Department of Physics, Northeastern University, Boston,
Massachusetts 02115, USA}}

\title{The 6-vertex model of hydrogen-bonded crystals with bond defects}

\begin{document}

\maketitle

\bigskip

\begin{abstract}
It is shown that the percolation model of hydrogen-bonded crystals, which is
a 6-vertex model with
bond defects, is completely equivalent with an 8-vertex model in an
external electric field.  Using this equivalence
we solve exactly a particular 6-vertex model with
bond defects. The general solution for the Bethe-like lattice is also
analyzed.

\end{abstract}

\vskip 1cm

\section{Introduction}
The 6-vertex model on a square lattice \cite{liebwu} describes
  hydrogen-bonded crystals in two dimensions.  Historically,
it was Slater \cite{slater} who first
considered the evaluation of the
residual entropy of ice, a hydrogen-bonded crystal, under the assumptions
that
(i) there is one hydrogen atom on each lattice edge, and (ii) there are always
two hydrogen atoms near, and away from, each lattice site (the ice rule).
Under these assumptions
 there are 6 possible hydrogen configurations at each site, and one is led to
a 6-vertex model.  The exact residual entropy of the ``square" ice,
i.e., ice on the square lattice,
 was obtained by Lieb \cite{lieb}, which gives  rise
to a numerical number surprisingly close to the experimental residual
entropy of real ice (in three dimensions).  The 6-vertex model is therefore
an accurate description of hydrogen-bonded crystals.

In real hydrogen-bonded crystals, however, there exist bonding defects
\cite{G82}.  One way through which  bond defects can occur is caused by
 the double-well potential seen by hydrogen atoms
between two lattice sites.  When two hydrogens
 occupy the two off-center potential wells along a given  lattice edge,
the assumption (i)
above is broken, albeit the ice rule (ii) is still intact.
This leads to the fact that one, two or zero hydrogen atoms can
be present on a lattice edge. Indeed, one of us \cite{hu}
has considered this possibility  in  a percolation model of supercooled
water. The same
model has later been considered by Attard and Batchelor \cite{ab} who
analyzed it using series analyses.  More recently, Attard \cite{attard}
reformulated the problem as a 14-vertex model with Bjerrum bond defects, and
analyzed the 14-vertex model using an independent bond approximation.
Here, using a somewhat different mapping, we establish the exact
equivalence of the 6-vertex model with
bond defects  with an 8-vertex model in an external field.  As a result, we
are able to analyse the exact solution in a particular parameter subspace.
We also discuss the general solution  of the 6-vertex model with
bond defects on the Bethe lattice.

\section{Equivalence with an 8-vertex model}
Consider a square lattice ${\cal L}$ of $N$ sites under periodic boundary
conditions   so that there are   $2N$ lattice edges.
The lattice is hydrogen-bonded with defects such that there can be
one, two or zero hydrogen atoms on each lattice edge.  As the hydrogen atoms
are placed off-center on the edges, we place two Ising spins
$\s, \s'$  on each lattice
edge such that $\s =1$ denotes that the site is  occupied by a hydrogen
and $\s=-1$ the site is empty.  However, the ice rule dictates that the
sum of the
4 Ising spins $\s_1,\s_2,\s_3,\s_4$ surrounding a square lattice sites
must vanish. There are altogether six ice-rule configurations as shown in Fig. 1.

\begin{figure}
\epsfxsize=70mm
\vbox to1in{\rule{0pt}{1in}}
\includegraphics{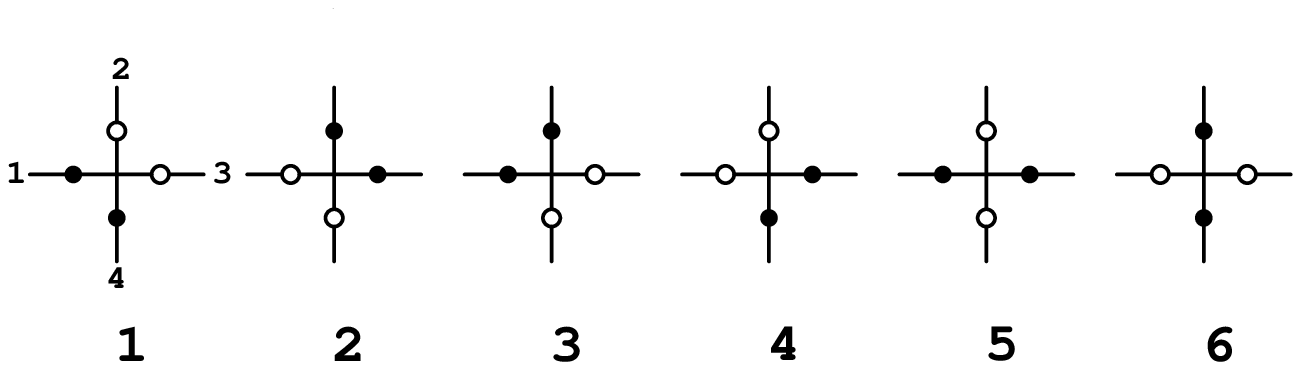}
\caption{The six ice-rule configurations.  Each solid circle
denotes an Ising spin $\sigma=1$ and each open circle a $\sigma=-1$.}
\end{figure}

More generally, we consider the  ice-rule model  with weights
 \be
\o_1=\o_2=a,\hskip 1cm \o_3=\o_4=b, \hskip 1cm \o_5=\o_6=c, \label{weights}
\ee
where $\o_i$ is the weight of the $i$th configuration shown in Fig. 1.
Denote the weights  (\ref{weights}) by $\o( \s_1,\s_2,\s_3,\s_4)$
where the subscripts are indexed as shown and $\o_1=\o(1,-1,-1,1)$, etc.
The weight $\o( \s_1,\s_2,\s_3,\s_4)$
 satisfies the ``spin reversal" symmetry
\be
\o( \s_1,\s_2,\s_3,\s_4) =\o( -\s_1,-\s_2,-\s_3,-\s_4)\label{symmetry}
\ee
and vanishes except for the six weights given in (\ref{weights}).
To each lattice edge containing  two spins $\s$
and $\s'$, introduce an edge factor $E(\s, \s')$ to reflect the effect
of bond defects.  Then, the partition function of interest is
\be
Z = \sum_{\s=\pm 1}\> \prod_{\rm vertices}\o( \s_1,\s_2,\s_3,\s_4)
 \prod_{\rm edges} E(\s, \s').  \label{part}
\ee
In the 6-vertex model without bond defects
we have $E(\s, \s')= (1 - \s \s')/2$ so that  there
is precisely one hydrogen on each lattice edge.
The model considered by Attard \cite{attard} is described by
\bea
E( \s, \s') &=& w_+, \hskip 1cm \s=\s'=1 \nonumber \\
             &=& w_-, \hskip 1cm \s=\s'=-1 \nonumber \\
              &=& 1, \hskip 1.35cm \s=-\s'.  \label{att}
\eea
The percolation model of \cite{hu} is equivalent to a special case of (\ref{att}) with
$w_+ = w_- = e^{2K}$ and
\be
E( \s, \s') =e^{K(\s\s'+1)}=e^{K}(\cosh{K})(1+z \s \s'), \label{e}
\ee
where $z=\tanh{K}$. For our purposes, we shall restrict our considerations to the
percolation model (\ref{e}).

Attard and Batchelor \cite{ab,attard} adopted an arrow representation for the
hydrogen configurations which, due to the occurrence of defects,
led to a 14-vertex model with Bjerrum defects.
A weak-graph transformation
\cite{nagle} is then carried out
for the 14-vertex model.   Here, we expand
 the partition function directly.
Substituting (\ref{e}) into (\ref{part}) and expanding the
second product over the edges of ${\cal L}$,
we obtain an expansion of $2^{2N}$ terms.  To each term
in the expansion we associate  a bond graph
by drawing  bonds on those edges corresponding to the $z$ factors
contained in the term.  This leads to a 16-vertex model
on ${\cal L}$. Besides an overall Boltzman factor $(e^K \cosh{K})^{2N}$,
the 16-vertex model has  vertex weights
\be
W = \sum_{\s_1\s_2\s_3\s_4} \biggl(\o( \s_1,\s_2,\s_3,\s_4) \prod (\sqrt z \s_i)\biggr),\label{w}
\ee
where the product is taken over those incident edges with bonds.
The symmetry relation (\ref{symmetry}) now implies that $W=0$ whenever there
are an odd number of incident bonds, and the 16-vertex model becomes an
8-vertex model.

\begin{figure}
\epsfxsize=70mm
\vbox to1in{\rule{0pt}{1in}}
\includegraphics{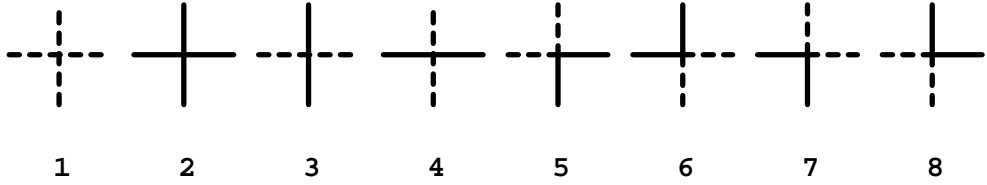}
\caption{The 8-vertex model configurations.}
\end{figure}

Using the bond configurations of the 8-vertex model shown in Fig. 2,
it is straightforward to deduce using (\ref{w}) the following vertex weights:
\bea
W_1&=& \sum \o( \s_1,\s_2,\s_3,\s_4) = 2(a+b+c) \nonumber \\
W_2&=& z^2\sum \s_1\s_2\s_3\s_4\>\o( \s_1,\s_2,\s_3,\s_4) = 2z^2(a+b+c) \nonumber \\
W_3&=& W_4 =z\sum \s_2\s_4\>\o( \s_1,\s_2,\s_3,\s_4) = 2z(-a-b+c) \nonumber \\
W_5&=& W_6 =z\sum \s_1\s_2\>\o( \s_1,\s_2,\s_3,\s_4) = 2z(-a+b-c) \nonumber \\
W_7&=& W_8 =z\sum \s_2\s_3\>\o( \s_1,\s_2,\s_3,\s_4) = 2z(a-b-c), \label{8weights}
\eea
where $W_i$ is the vertex weight of the $i$th vertex and
we have used the fact that in the non-vanishing $\o$'s we have
$\s_1\s_2\s_3\s_4=1$.  Now, in an 8-vertex model  configurations,
vertices 3 and 4, 5 and 6, and 7 and 8, always occur in pairs and/or in even
numbers.
Therefore we can  conveniently replace the relevant weights by their absolute
values and arrive at, after dividing all weights by a common factor $z$,
\bea
W_1&=& z^{-1}(a+b+c) \nonumber \\
W_2&=& z(a+b+c) \nonumber \\
W_3&=& W_4 =|-a-b+c| \nonumber \\
W_5&=& W_6 = |-a+b-c| \nonumber \\
W_7&=& W_8  = |a-b-c|. \label{8weights1}
\eea
The vertex weights (\ref{8weights1}) describes an 8-vertex model in an
external electric field $h=(\ln z)/2$ in both the vertical and horizontal
directions \cite{liebwu}.

More generally, if we allow different values of $w_+=w_-=w_i, i=1,2$ in (\ref{att}) and
(\ref{e}) for the horizontal and vertical edges respectively,
and write $h=(\ln z_1)/2$ and $v=(\ln z_2)/2$,
where $z_i=(w_i-1)/(w_i+1), i=1,2$, then one arrives at an 8-vertex model with
weights
\bea
W_1&=& e^{h+v}(a+b+c) \nonumber \\
W_2&=& e^{-h-v}(a+b+c) \nonumber \\
W_3&=& e^{h-v}|-a-b+c| \nonumber \\
W_4&=& e^{v-h}|-a-b+c| \nonumber \\
W_5&=& W_6 = |-a+b-c| \nonumber \\
W_7&=& W_8  = |a-b-c|. \label{8weights2}
\eea
In ensuing discussions we shall consider the general model (9).

\section{The free-fermion solution}

The free-fermion model \cite{fw70} is defined as a particular case of the
eight-vertex model in which the vertex weights satisfy the relation
\be
W_1W_2 + W_3W_4 = W_5W_6 + W_7W_8,
\ee
a condition  equivalent to the consideration
of a noninteracting many-fermion system in an $S$-matrix formulation of the
8-vertex model \cite{hg60}.
In the present case the free-fermion condition (10) is
satisfied when either $a=0$ or $b=0$ . Without the loss of generality, we
 consider $a=0$ and $b>c$.

The closed expression for the free energy of the free-fermion model is
well-known \cite{fw70}  and after some algebraic manipulation using results
of \cite{fw70}, we obtain
\be
-\beta f=\lim_{N \to \infty}\frac{1}{N}\ln{Z}=
\ln{(b+c)}+\frac{1}{4 \pi}\int_0^{2 \pi}d{\phi}\ln{(A+Q^{1/2})},
\label{freeenergy}
\ee
where
\bea
Q&=&[\sinh{(2v+2h)}+k^2\sinh{(2v-2h)}+2k\cosh{2v}\cos{\phi}]^2+
4 k^2{\sin{\phi}}^2,
\nonumber \\
A&=&\cosh{(2v+2h)}+k^2\cosh{(2v-2h)}+2k\sinh{2v}\cos{\phi}
\label{coeff}
\eea
with $k=(b-c)(b+c)$.

The critical condition of the free-fermion model is given by \cite{fw70}
\be
W_1 + W_2 + W_3 + W_4 = 2 \max{\{W_1,W_2,W_3,W_4\}},
\ee
where
\bea
W_1&=& e^{h+v}(b+c) \nonumber \\
W_2&=& e^{-h-v}(b+c) \nonumber \\
W_3&=& e^{h-v}(b-c) \nonumber \\
W_4&=& e^{v-h}(b-c). \label{8weights3}
\eea
Thus, the $h$-$v$ plane is divided into four regions depending
on which vertex $1$, $2$, $3$ or $4$ has the largest weight.
Denoting the four regions by $I$, $II$, $III$ and $IV$ respectively, as
shown in Fig. 3, the critical condition (13) can be rewritten as
\begin{figure}
\epsfxsize=70mm
\vbox to2.5in{\rule{0pt}{2.5in}}
\includegraphics{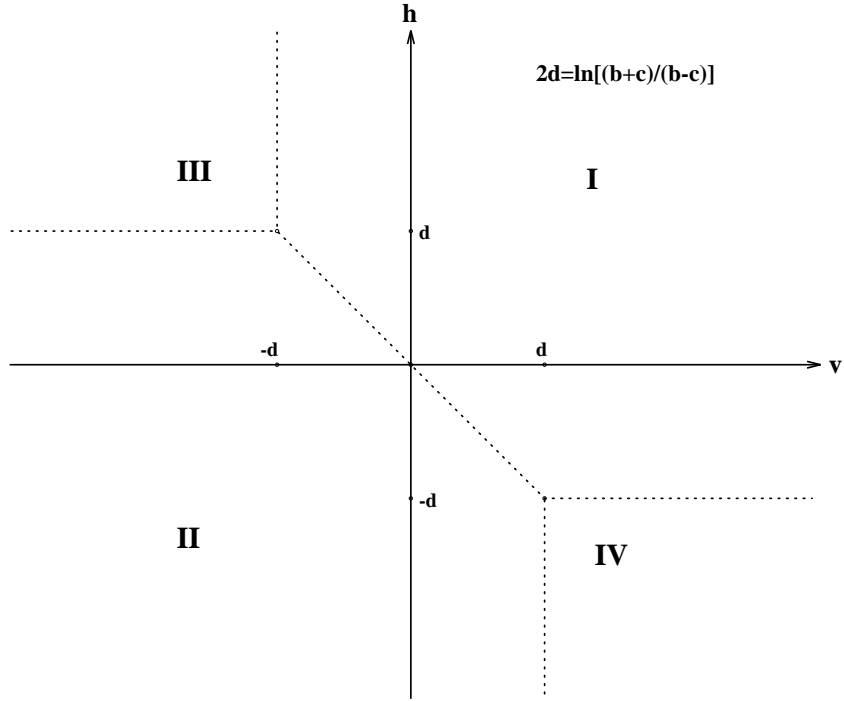}
\caption{The four regions in the plane $h$-$v$
denoted by $I$, $II$, $III$ and $IV$ are, respectively, the regions where
vertex 1, 2, 3 or 4 has lowest energy.}
\end{figure}

\bea
\frac{b}{c}&=&\frac{\tanh{v}+e^{2 h}}{1-e^{2 h} \tanh{v}}, \qquad \qquad \mbox{region} \quad I,
\nonumber \\
\frac{b}{c}&=&\frac{1-e^{2 h}\tanh{v}}{\tanh{v}+e^{2 h}}, \qquad \qquad \mbox{region} \quad II,
\nonumber \\
\qquad \frac{b}{c}&=&\frac{\tanh{v}-e^{2 h}}{1+e^{2 h} \tanh{v}}, \qquad \qquad \mbox{region} \quad III,
\nonumber \\
\qquad \frac{b}{c}&=&\frac{1+e^{2 h}\tanh{v}}{\tanh{v}-e^{2 h}}, \qquad \qquad \mbox{region} \quad IV.
\label{critical}
\eea
The critical condition (15) is plotted in Fig. $4a-4d$ for four specific
values of $b/c$. Generally, the free energy exhibits a logarithmic singularity at the phase boundaries
with exponents $\alpha={\alpha}'=0$ \cite{fw70}. When $b=c$ for which $Q$ is a
complete square, however, we have
\bea
-\beta f &=&\max{\{\ln{W_1},\ln{W_2}\}}  \nonumber \\
&=& \ln{(2b)}+|h+v| \label{complete}
\eea
and the phase boundary $h+v=0$ separates the two frozen states $W_1$ and $W_2$.

\begin{figure}
\epsfxsize=70mm
\vbox to3in{\rule{0pt}{3in}}
\includegraphics{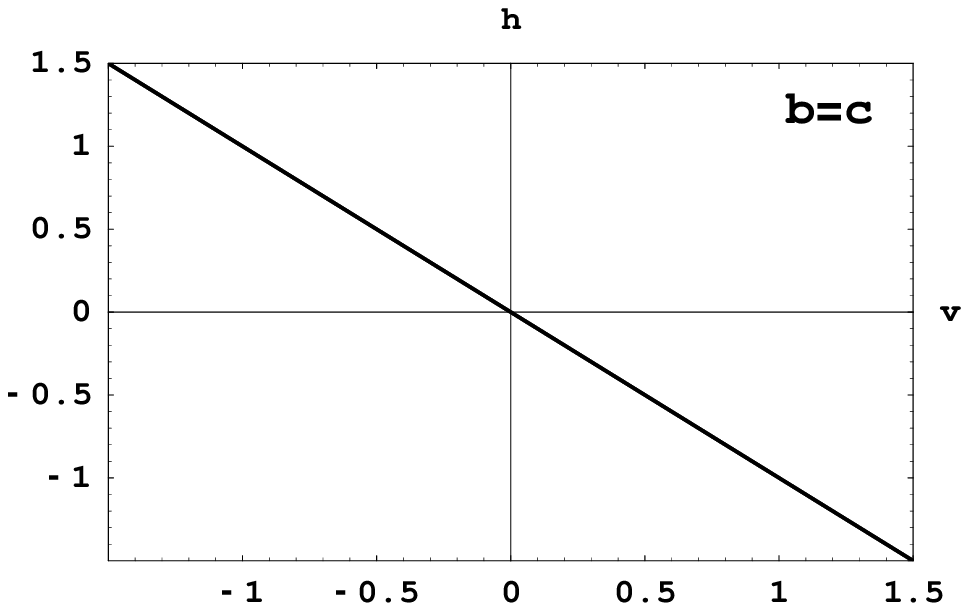}
\includegraphics{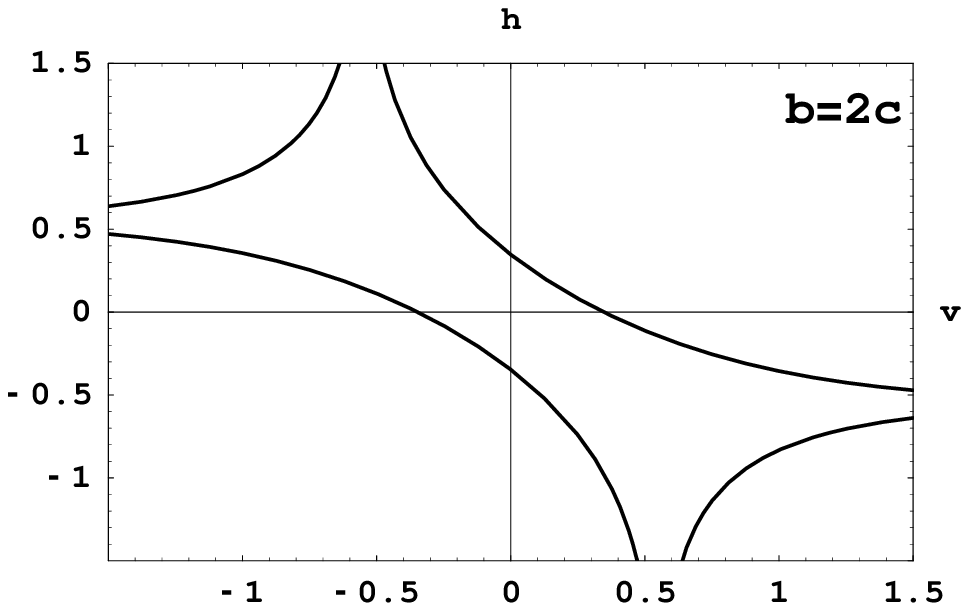}
\includegraphics{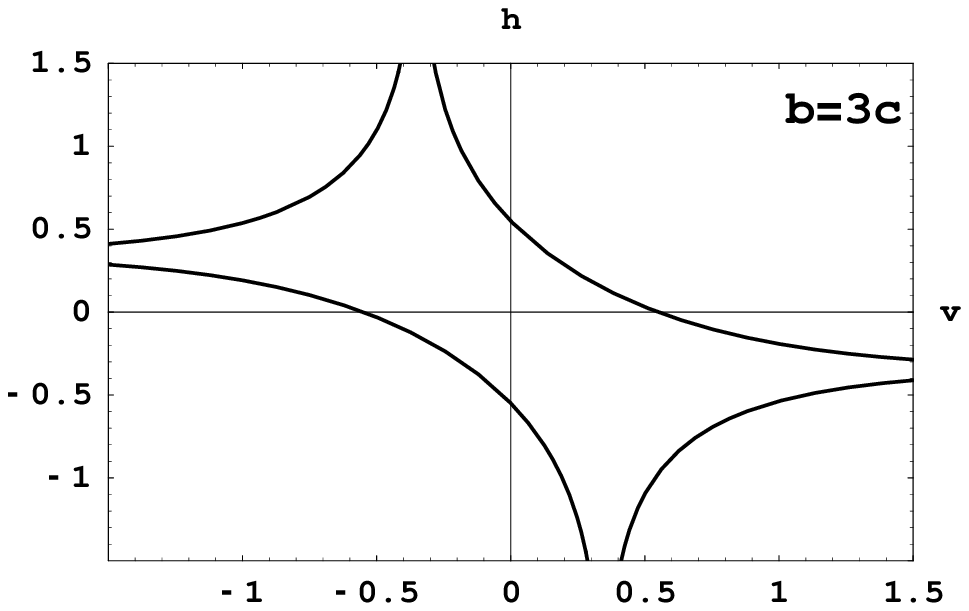}
\includegraphics{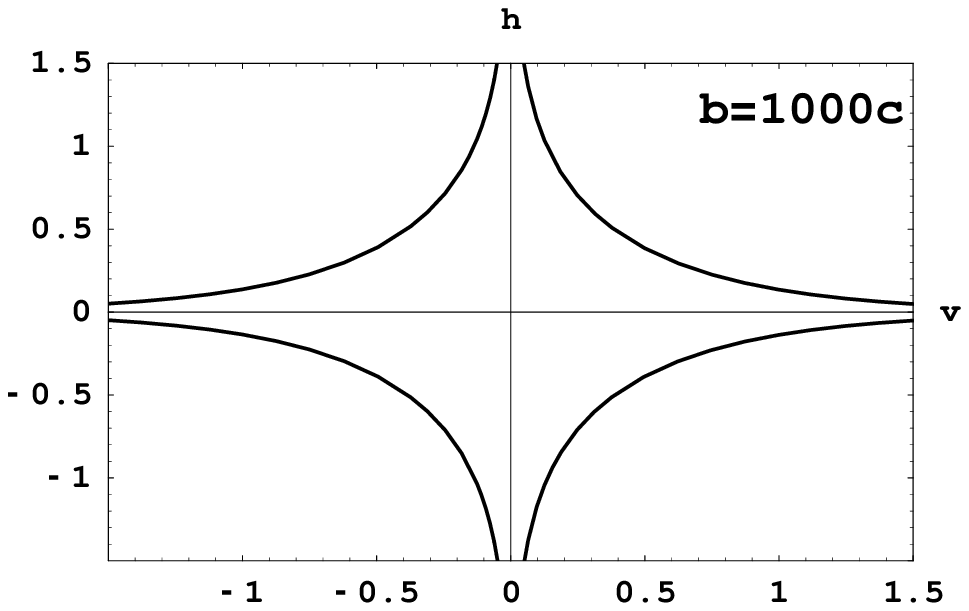}
\caption{Phase boundaries of the free-fermion model.}
\end{figure}

\section{The Bethe-like lattice}

Now, we consider the 6-vertex model with bond defects on a
Bethe-like lattice with plaquettes as shown in Fig. 5.
The study of systems in the Bethe-like
lattices is an alternative approach to the usual mean-field theory.
The main features of the model under investigation will be obtained by
studying the properties of the free energy.

A free energy in a region deep inside a Bethe-like lattice must be carefully
defined. It cannot be
obtained by directly evaluating the logarithm of the partition function in
which the contribution from the outside of this region is not negligible
and as result the system exhibits an unusual type of phase transitions
without long-range order \cite{e74,mz74,ww76}. Recently, a method for the
surface independent free energy calculation is presented \cite{g95,aaio}.
The free energy $f_{\Box}$ per plaquette of our model is expressed as
\be
-\beta f_{\Box}=\lim_{n \to \infty}\frac{1}{2}\left(\ln{Z_n}-3 \ln{Z_{n-1}}\right)
\label{fener}
\ee
where $Z_n$ and $Z_{n-1}$ is the partition functions of the 6-vertex model
with bond defect on the Bethe-like lattice consist of $n$ and $n-1$
generations respectively.
\begin{figure}
\epsfxsize=70mm
\vbox to2in{\rule{0pt}{2in}}
\includegraphics{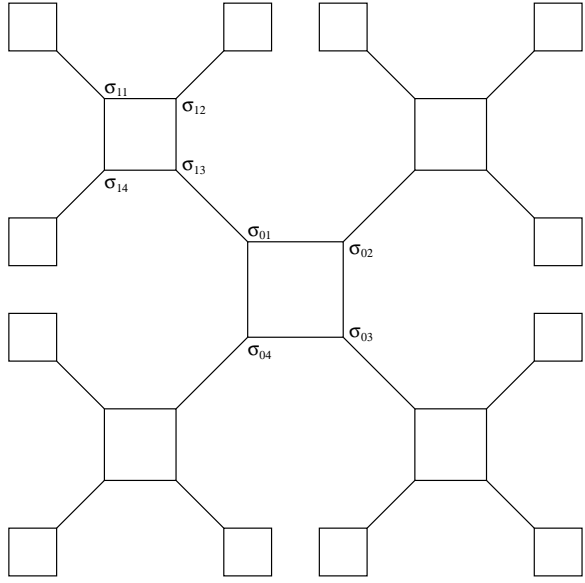}
\caption{The Bethe-like lattice.}
\end{figure}

The calculation on the Bethe-like lattice is based on a recursion method \cite{ih98}.
When the tree is cut at the central plaquette, it is separate into 4 branches,
each of which
contains 3 branches.  Then the partition function of interest (\ref{part})
can be written as follows:
\be
Z_n = \left(\frac{\o + 1}{2} \right)^{N_b^{(n)}} \sum_{\{\s_{0i}\}} \o(\s_{01},
\s_{02},\s_{03},\s_{04})g_n(\s_{01})g_n(\s_{02})g_n(\s_{03})g_n(\s_{04}),
\label{zbethe}
\ee
where $N_b^{(n)} = 2(3^n-1)$ is the number of bonds, $n$ is the  number of
generations and $g_n(\s_{0i})$  is in fact the partition
function of a branch nearest to the $0i$ site.

Each branch, in turn, can be cut along any site of the first generation.
then the expressions for $g_n(\s_{0i})$ can therefore be written in
the form:
\be
g_{n+1}(\s_{01})= \sum_{\{\s_{1i}\}}(1+z \s_{01} \s_{13}) \o(\s_{11},
\s_{12},\s_{13},\s_{14})g_n(\s_{11})
g_n(\s_{12})g_n(\s_{14})
\label{recrel}
\ee
After dividing $g_n(-)$ by $g_n(+)$, we obtain a recursion relation for
$x_n=g_n(-)/g_n(+)$. Let us  consider the case when series of solution of
recursion relation converge to a stable point at $n \to \infty$, namely,
$$
\lim_{n \to \infty}x_n = x.
$$
We obtain the following equation:
\be
x = \frac{(1-z) x^2 + (1+z) x}
{(1+z) x^2 + (1-z) x}.
\label{eqstat}
\ee
We are now in a position to compute the free energy per plaquette of our
model. Using Eqs. (\ref{fener}), (\ref{zbethe}), (\ref{recrel}) and
(\ref{eqstat}), the expression for the free energy functional can be written
as
\bea
-\beta f_{\Box}&=&\lim_{n \to \infty}\frac{1}{2}\left(N_b^{(n)}-
3 N_b^{(n-1)}\right)\ln{\frac{\o +1}{2}} \label{free} \\
&+&\lim_{n \to \infty}\frac{1}{2}\left[\ln{\Phi(x_{(n-1)})}+
\ln{\Psi(x_{(n)})}-3 \ln{\Psi(x_{(n-1)})}\right] \nonumber
\eea
where
\be
\Psi(x)=2(a+b+c)x^2 \quad \mbox{and} \quad
\Phi(x) =(a+b+c)^4 x^4 [(1+z) x + 1-z]^4  \nonumber
\ee
It easy to see that $N_b^{(n)}-3 N_b^{(n-1)}=4$ for all $n$.
Thus the free energy per plaquette can be finally written as
\be
-\beta f_{\Box} = \ln{\frac{(\o+1)^2}{2}(a+b+c)}
+\ln{\frac{(1+z) x + 1-z}{2}}.
\label{energy}
\ee
Together with the expression (\ref{eqstat}) for $x$ it gives the
free energy per plaquette of the 6-vertex model with bond defects.

The equation of state (\ref{eqstat}) always has $x=1$ as
fixed point solution. In this case the free energy per plaquette is
\be
-\beta f_{\Box}=\ln{\frac{(w+1)^2}{2}(a+b+c)}
\ee
In the case of $a=b=c=1$, we recover the result obtained by Attard and
Batchelor \cite{ab} for the six-vertex model with bond defect in the
mean-field (independent vertex) framework,
$$
-\beta f_{\Box}=\ln{\frac{3}{2}(w+1)^2},
$$
which reduces to Pauling's estimate when $w=0$ \cite{p35}.
\begin{figure}
\epsfxsize=70mm
\vbox to2.5in{\rule{0pt}{2.5in}}
\includegraphics{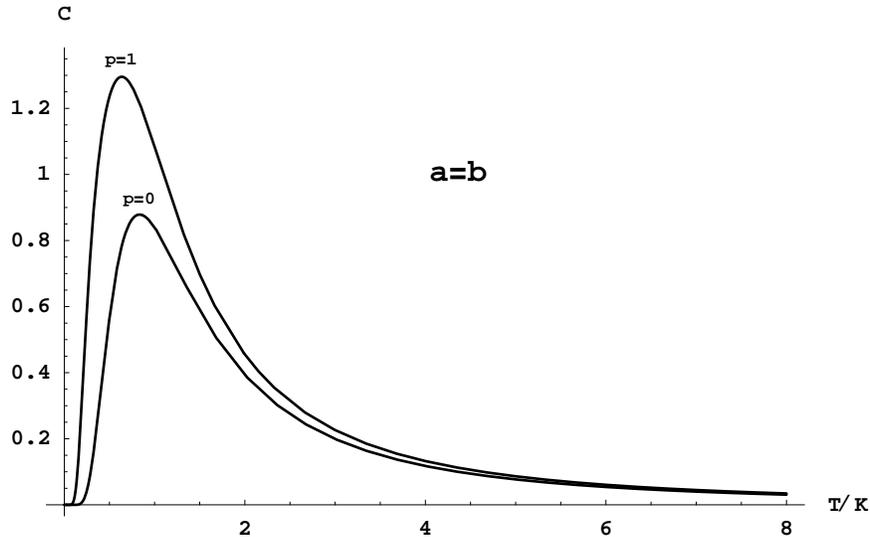}
\caption{ The specific heat for the Bethe-like lattice of Fig. 5 as
function of $T/K$ for $a=b$ and for $p=(\varepsilon_a - \varepsilon_c)/K =
0, 1$.}
\end{figure}

For the specific heat, we obtain
\bea
C&=&\frac{d}{d T}\left[T^2\frac{d}{d T}(-\beta f_{\Box}
)\right] \nonumber \\
&=&\frac{2w}{(w+1)^2}(\ln{w})^2+\frac{a b(\ln{a/b})^2
+a c(\ln{a/c})^2+b c(\ln{b/c})^2}{(a+b+c)^2},
\label{specificheat}
\eea
or, explicitly,
\be
C=\frac{1}{T^2}
\frac{2 K^2}{(\cosh{K/T})^2}+
\frac
{
{\varepsilon}_{ab}^2 e^{{\varepsilon}_{ab}/T}+
{\varepsilon}_{ac}^2 e^{{\varepsilon}_{ac}/T}+
({\varepsilon}_{ab}-{\varepsilon}_{ac})^2 e^{({\varepsilon}_{ab}+{\varepsilon}_{ac})/T}
}
{
T^2(1+e^{{\varepsilon}_{ab}/T}+e^{{\varepsilon}_{ac}/T}
)^2
}
\label{specificheat1}
\ee
where ${\varepsilon}_{ab}={\varepsilon}_a -{\varepsilon}_b,  \quad
{\varepsilon}_{ac}={\varepsilon}_a -{\varepsilon}_c$ and

$$
a=e^{-{\varepsilon}_a/T}, \quad \quad
 b=e^{-{\varepsilon}_b/T},\quad \quad c=e^{-{\varepsilon}_c/T}.
$$
The specific heat versus $T/K$ for $a=b$ is plotted in Fig. 6.

\section{Summary and discussion}

In this paper we have considered a
6-vertex model of hydrogen-bonded crystals with bond defects.
We have established the exact equivalence of the 6-vertex model with bond
defects with an 8-vertex model in an external electric field.
Using this equivalence we solve exactly our model in the free-fermion
subspace. We also obtain the exact solution of the 6-vertex model with bond
defects on a Bethe-like lattice.

In \cite{hu}, one of us has used the percolation representation of the
hydrogen-bonded model to argue that the specific heat of the system will
increase as the temperature decreases from very high temperature. Figure 6
indeed shows such behavior. However, the specific heat in Fig. 6 does not
diverge. It is of interest to calculate the specific heat for the square or
diamond lattices by Monte Carlo method. We expect to find singular behavior
of specific heat in such systems.

\vskip 8 mm
\centerline{ACKNOWLEDGMENTS}

\medskip
This work was partly supported  by the National Science Council of
the Republic of China (Taiwan) under grant number NSC 89-2112-M-001-005.
The work of FYW is supported in part by the National Science Foundation
grant DMR-9614170. FYW thanks C. K. Hu for the hospitality at the Academia
Sinica and T. K. Lee for the hospitality at the National Center for
Theoretical Sciences where this work is completed.

\end{document}